\documentclass[onecolumn,a4paper,fleqn,usenatbib,linenumbers]{mnras}
\usepackage{amsmath,amssymb,CJK,soul}
\usepackage{longtable}
\usepackage{hyperref}
\usepackage{longtable}
\usepackage{threeparttablex} 
\usepackage{newtxtext,newtxmath}
\usepackage[T1]{fontenc}
\usepackage{ae,aecompl}

\usepackage{graphicx}	
\usepackage{amsmath}	
\usepackage{amssymb}	

\title[Constraints on the Spin-Pole, Jet Morphology and Rotation of 2I]{Constraints on the Spin-Pole Orientation, Jet Morphology and Rotation of Interstellar Comet 2I/Borisov with Deep \textit{HST} Imaging}

\author[B.T. Bolin et al.]{
Bryce T. Bolin$^{1,2}$\thanks{E-mail: bbolin@caltech.edu (BTB)}
and Carey M. Lisse$^{3}$
\\
$^{1}$Division of Physics, Mathematics and Astronomy, California Institute of Technology, Pasadena, CA 91125, U.S.A.\\
$^{2}$IPAC, Mail Code 100-22, Caltech, 1200 E. California Blvd., Pasadena, CA 91125, U.S.A.\\
$^{3}$Johns Hopkins University Applied Physics Laboratory, Laurel, MD 20723, U.S.A.
}

\date{Accepted 23 July 2020. Received 3 June 2020; in original form 5 May 2020}

\pubyear{2020}

\begin{document}
\label{firstpage}
\pagerange{\pageref{firstpage}--\pageref{lastpage}}
\maketitle

\begin{abstract}
We present high resolution, deep imaging of interstellar comet 2I/Borisov taken with the \textit{Hubble Space Telescope}/Wide Field Camera 3 (\textit{HST}/WFC3) on 2019 December 8 UTC and 2020 January 27 UTC (HST GO 16040, PI Bolin) before and after its perihelion passage in combination with \textit{HST}/WFC3 images taken on 2019 October 12 UTC and 2019 November 16 UTC (HST GO/DD 16009, PI Jewitt) before its outburst and fragmentation of March 2020, thus observing the comet in a relatively undisrupted state. We locate 1-2\arcsec~long (2,000 - 3,000 km projected length) jet-like structures near the optocenter of 2I that appear to change position angles from epoch to epoch. With the assumption that the jet is located near the rotational pole supported by its stationary appearance on $\sim$10-100 h time frames in \textit{HST} images, we determine that 2I's pole points near $\alpha$ = 322$\pm$10$^\circ$, $\delta$ = 37$\pm$10$^\circ$ ($\lambda$ = 341$^\circ$, $\beta$ = 48$^\circ$) and may be in a simple rotation state. Additionally, we find evidence for possible periodicity in the \textit{HST} time-series lightcurve on the time scale of $\sim$5.3 h with a small amplitude of $\sim$0.05 mag implying a lower limit on its $b/a$ ratio of $\sim$1.5 unlike the large $\sim$2 mag lightcurve observed for 1I/`Oumuamua. However, these small lightcurve variations may not be the result of the rotation of 2I's nucleus due to its dust-dominated light-scattering cross-section. Therefore, uniquely constraining the pre-Solar System encounter, pre-outburst rotation state of 2I may not be possible even with the resolution and sensitivity provided by \textit{HST} observations.
\end{abstract}

\section{Introduction}
The second interstellar object, 2I/Borisov (hereafter 2I), has had a morphological appearance similar to Solar System comets ever since its discovery \citep[][]{Guzik2019a, Bolin2020aa}. Cometary gas measurements have revealed CO \citep[][]{Bodewits2020,Cordiner2020}, and [O I] resulting from photodissociation of H$_2$O \citep[][]{McKay2019,Crovisier2019}, and the typical visible-wavelength daughter species \citep[e.g., CN and C2][]{Fitzsimmons2019, Opitom2019,Bannister2020}. The comet seems to have a high CO abundance but may be be carbon-chain depleted \citep[][]{AHearn1995,Kareta2019}.

The color and activity of 2I have been revealed to be similar to that of Solar System comets. Visible to near-infrared photometry and spectra of the comet are reddish to neutral suggesting the presence of refractory organics and water-ice common in comets \citep[][]{deLeon2020,Bolin2020aa,Yang2020}. Additionally, its activity has been revealed by long-term lightcurve trending to be driven by super-volatiles such as CO/CO$_2$ far from the Sun, with the addition of H$_2$O-driven activity as the comet crossed inside 3-4 au on its way to a $\sim$2 au perihelion distance \citep[][]{Bolin2020aa,Ye2019b, Cremonese2020}. Constraints on 2I's size from gas production rates \citep[][]{Fitzsimmons2019}, statistical arguments \citep[][]{Jewitt2019}, high resolution adaptive optics imaging from ground-based facilities \citep[][]{Bolin2020aa} suggest a nucleus $\sim$ km-scale diameter. However, direct detection of the nucleus has not yet been possible due to the total brightness of the coma obscuring the nucleus \citep[][]{Jewitt2019}. 

In this paper, we present high resolution and deep stack optical observations made with the \textit{Hubble Space Telescope} to study the coma structure of 2I. In particular, our observations are relevant in studying the comet before it underwent a 0.7 magnitude outburst on 2020 March 7 UTC, 2I \citep[][]{Drahus2020Atel,Jehin2020CBET}. Subsequent HST observations in late 2020 March revealed that 2I was undergoing fragmentation in possible connection to the outburst \citep[][]{Bolin2020Atel, Jewitt2020aadfasd}. Regardless of the cause of its outburst, we present these observations as indicative of 2I in its pre-outburst state before fragmentation may have had an affect on its physical properties and rotation.

\section{Observations}
The \textit{Hubble Space Telescope} (\textit{HST}) was used to observe 2I with General Observer's (GO) time on 2019 December 8 UTC and 2020 January 27 UTC \citep[HST GO 16040, PI][]{BolinHST2019b}. Two orbits were used to observe 2I on 2019 December 8 UTC and two more orbits were used on 2020 January 27 UTC. During each orbit on 2019 December 8 UTC, one 404 s F689M filter exposure and one 404 s F845M filter exposure was obtained with the UVIS2 array of the WFC3/UVIS camera \citep[][]{Dressel2012} for a total of 4 exposures and 1616 s integration time over both orbits. For the two orbit observations of 2I on 2020 January 27 UTC, two 386 s exposures were obtained in the F689M and in the F845M filter for a total exposure time of 1544 s. The F689M filter has a central wavelength of 688.32 nm with a FWHM bandpass of 70.76 nm and the F845M filter has a central wavelength of 846.40 nm with a FWHM bandpass of 88.07 nm \citep[][]{Deustua2017}. The comet was observed on two back-to-back orbits and was tracked non-sidereally.

As already described in \citep[][]{Manzini2020,Kim2020}, \textit{HST} was used to observe 2I with director's discretionary time on 2019 October 12 UTC \citep[HST GO/DD 16009,][]{Jewitt2019hst}. Up to seven orbits were awarded for the director's discretionary time, with four orbits were used on 2019 October 12 UTC to observe 2I. During each orbit on 2019 October 12 UTC, six 260 s F350LP filter exposures were obtained with the 2K subarray of the WFC3/UVIS camera for a total of 24 exposures and 6240 s integration time. The F350LP filter has a central wavelength of 581.95 nm with a FWHM bandpass of 489.26 nm \citep[][]{Deustua2017}. The comet was observed on four back-to-back orbits and was tracked non-sidereally at its rate of sky motion. Additional observations of 2I were conducted with one orbit of \textit{HST} on 2019 November 16 UTC. Six 230 s exposures were taken with the F350LP filter tracking on its rate of motion, though only four of the 230 s exposures were used from November 16 UTC because the coma of 2I significantly overlapped with an extended background galaxy in two of the 230 s exposures resulting in a total exposure time of 920 s. The viewing geometry of 2I during all of the \textit{HST} observations described here are available in Table~\ref{t:hstobs}. The resolution element in all imaging data presented here is 0.04\arcsec/pixel.

\begin{table}
\centering
\caption{WFC3/UVIS F350LP equivalent V-band photometry and viewing geometry.}
\label{t:hstobs}
\begin{tabular}{lclclclclclclc|c|c|c|}
\hline
\hline
Date$^1$& $H^2$ & $\sigma_H^3$ & $r_h^4$ & $\Delta^5$ & $\alpha^6$ & $\delta_{\earth}^7$&$b^8$& PA$^9$ & PID$^{10}$\\(UTC) & & & (au)& (au)&$(^\circ)$&$(^\circ)$&$(^\circ)$&$(^\circ)$&
\\ \hline
2019 Oct 12 & 16.59& 0.03& 2.38 & 2.79 & 20.26 & -14.23& 46.77& 180&GO/DD 16009\\
2019 Nov 16 & 16.57 & 0.03& 2.07 & 2.21 & 26.49 & -17.55& 48.87& 210&GO/DD 16009\\
2019 Dec 8 &16.54 & 0.04& 2.01 & 1.99 & 28.49 & -15.88& 40.58& 235&GO 16040\\
2020 Jan 27 &16.59 & 0.04& 2.29 & 2.04 & 25.51 & -0.66& 10.30& 250&GO 16040\\
\end{tabular}
\\(1) Date of observation, (2) H magnitude using a 0.2\arcsec radius aperture and 0.2-0.8\arcsec sky and outer coma subtraction annulus and Eq.~\ref{eqn.brightness}, (3) V magnitude uncertainty, (4) heliocentric distance, (5) geocentric distance, (6) phase angle, (7) Earth and target orbital plane angle, (8) galactic latitude of observations, (9) position angle of jet, (10) \textit{HST} program ID.
\end{table}

\section{Results}
\label{sec:results}
\subsection{Morphology and Jets}
\label{sec:jets}

\begin{figure}
\centering
\includegraphics[scale=0.51]{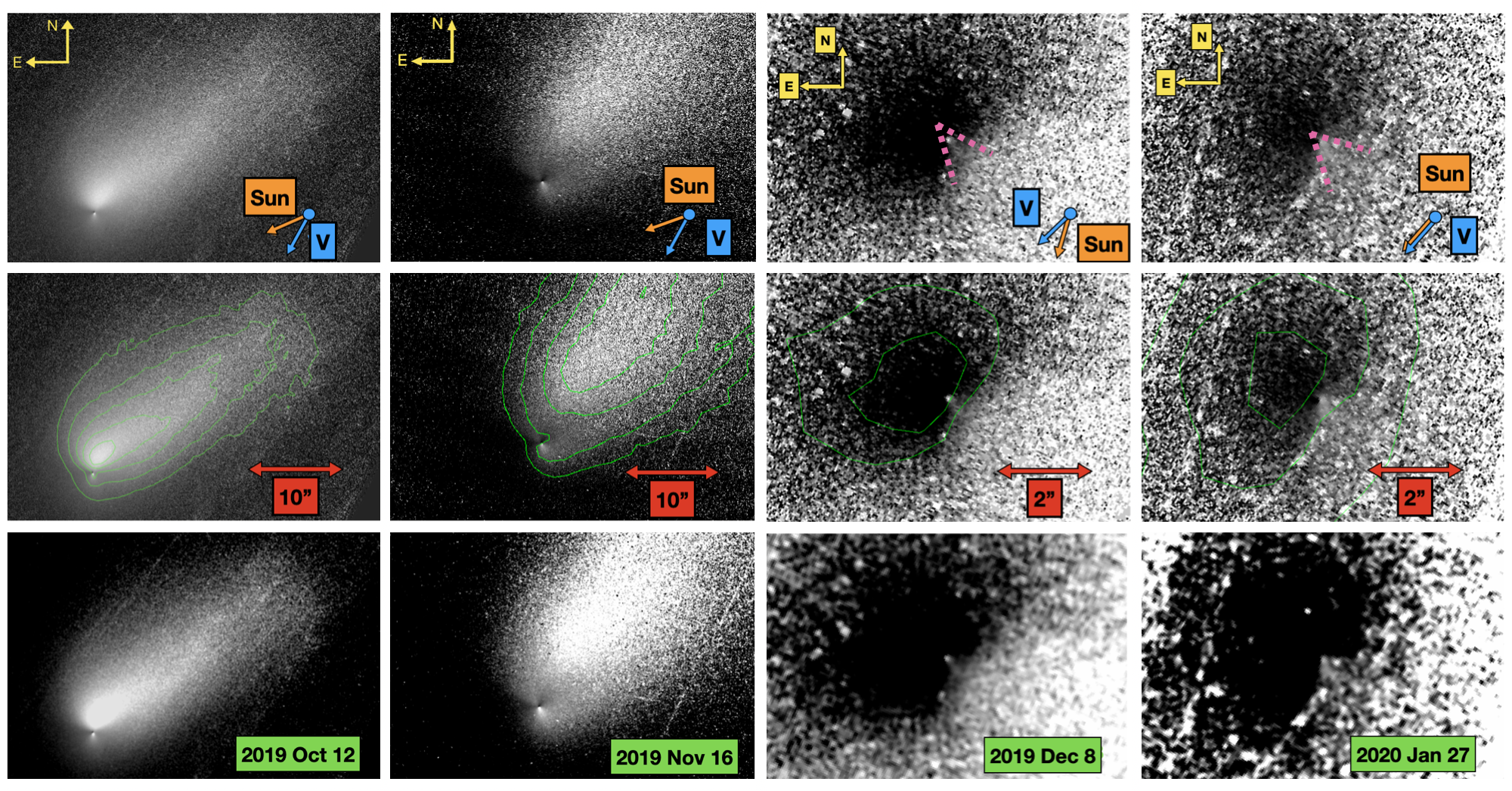}
\caption{Mosaic of composite images of 2I taken with \textit{HST} on 2019 October 12 UTC and 2019 November 16 UTC  using the F350LP filter \citep[HST GO/DD 16009, ][]{JewittHST2019a,Kim2020} and on 2019 December 8 UTC and 2020 January 27 UTC using the F689M and F845M filters \citep[HST GO 16040, ][]{BolinHST2019b}. Left column, top panel: 2019 October 12 UTC median stack of 24 x 260 s F350LP exposures taken on subsequent orbits with the radial profile of the coma removed The pixel scale is 0.04\arcsec/pixel. Left column, center panel: contour plot of the surface brightness profile of 2I over-plotted on the composite WFC3/UVIS image. 10\arcsec~is equivalent to $\sim$20,000 km at the geocentric distance of the comet of 2.79 au on 2019 October 12 UTC. Left column, bottom panel: Gaussian-smoothed version of the coma profile removed \textit{HST} composite image. Second left column: the same as the left panel, 2019 November 16 UTC median stack of 4 x 230 s F350LP exposures with the radial profile of the coma removed. 10\arcsec~is equivalent to $\sim$16,000 km at the geocentric distance of the comet of 2.21 au on 2019 November 16 UTC. Third column:  2019 December 8 UTC median stack of 2 x 404 s F689M and 2 x 404 s F845M exposures with the radial profile of the coma removed. 2\arcsec~is equivalent to $\sim$2,900 km at the geocentric distance of the comet of 1.99 au on 2019 December 8 UTC. The pink dotted wedges in the top panel highlights the location of the jet. Fourth column: 2020 January 27 UTC median stack of 2 x 386 s F689M and 2 x 386 s F845M exposures with the radial profile of the coma removed. 2\arcsec~is equivalent to $\sim$3,000 km at the geocentric distance of the comet of 2.04 au on 2020 January 27 UTC. The artifacts on the west side of the image is from incomplete removal of background stars at the low $\sim$10.3$^\circ$ galactic latitude of the observations.}
\label{fig.hstmosaic1}
\end{figure}

The two 404 s F689M filter exposures and two 404 s F845M filter obtained on 2019 December 8 UTC under HST GO 16040 were aligned and median stacked into a single composite image with an equivalent exposure time of 1616 s. Cosmic ray removal was done within the vicinity of the comet by interpolating any regions of the chip affected by cosmic rays with the average pixel values pf regions surrounding the cosmic ray strikes. The composite stacks were enhanced by normalizing the radial profile of the coma originating from the optocenter in the images. The enhanced coma removed images from 2019 December 8 UTC are presented in the third column of Fig.~\ref{fig.hstmosaic1}. The fourth column of Fig.~\ref{fig.hstmosaic1} presents a composite 1544 s stack of two F689M filter images and two F845M images.The streaked background in the 2020 January 27 UTC composite images is due to imperfect removal of dense background star fields at the 10.3$^\circ$ galactic latitude the images were taken. We combine our HST GO 16040 images with images taken by HST GO/DD 16009 \citep[PI][]{JewittHST2019a} on 2019 October 12 UTC and 2019 November 16 UTC. The 24 F350LP images obtained were median binned and cosmic ray removed identically to the F689M and F845M images. The enhanced F350LP coma removed images are also presented in the left-most column on Fig.~\ref{fig.hstmosaic1}. 

In the \textit{HST} observations on 2019 October 12 UTC, a clear tail is present in excess of 30\arcsec, as also seen in the ground-based observations \citep[][]{Jewitt2019, Bolin2020aa}, with a position angle opposite of the orbital velocity indicating the presence of 100 micron-sized dust grains\citep[][]{Jewitt2019hst,Kim2020}. However, the position angle of the tail may also be compatible with grains moving in the anti-solar direction; the two cases are difficult to distinguish due to projection effects at the modest orbital plane angle between \textit{HST} and the comet at the time of the observations. The enhanced coma removed images for the 2019 November 16 data are presented in the column second from the left on Fig.~\ref{fig.hstmosaic1}.The comet had a heliocentric distance of 2.07 au, a geocentric distance of 2.21 au, a phase angle of 26.49$^{\circ}$ and an orbital plane angle of $-$17.55$^{\circ}$ on 2019 November 16 UTC.

For the data taken on 2020 October 12 UTC presented in the center-left panel of Fig.~\ref{fig.hstmosaic1}, contours showing the structure of the tail and jet are over-plotted on the radial profile normalized composite image show an enhancement in surface brightness of the coma near its center. A smoothed-version of the radial profile normalized image is shown in the bottom left panel revealing an azimuthally asymmetric surface brightness profile within 5\arcsec~ of the center of the coma at position angles 0$^{\circ}$ and 180$^{\circ}$ \citep[][]{Jewitt2019hst, Manzini2020, Kim2020}. The thin southern surface brightness enhancement at a position angle of $\sim$180$^{\circ}$, more easily seen in the left panel's bottom image of Fig.~\ref{fig.hstmosaic1} that has been smoothed with a Gaussian filter to enhance low surface brightness features, may be interpreted as a localized jet, as seen in some Solar System comets \citep[e.g.,][]{Farnham2009,Knight2013,Bodewits2018}. The position angle of the jet has been estimated by the location of the peak brightness in the jet's radial profile drawn from the center of the nucleus in azimuthal coordinates with an origin of 0$^\circ$ pointing north \citep[e.g.,][]{Farnham2002}. There was no indication that the position angle of the jet showed significant variation or curvature as a function of radial distance from the coma even after the application of radial coma removal and smoothing contrast enhancement techniques \citep[][]{Schleicher2004}.

The F350LP data from 2019 November 16 UTC were also median stacked with a total equivalent exposure time of 920 s with the coma's radial profile normalized as seen in the second left column of Fig. ~\ref{fig.hstmosaic1}. Similarly, both jets are seen in the 2019 October 12 data appear in the 2019 November 16 UTC data in the third column of Fig. ~\ref{fig.hstmosaic1} with position angles of $\sim$30$^{\circ}$ and $\sim$210$^{\circ}$ as seen in the second column of Fig. ~\ref{fig.hstmosaic1}. The south-pointing jet is seen again in the 2019 December 8 UTC F689M + F845M data with a position angle of $\sim$235$^{\circ}$ as seen in the third column of Fig. ~\ref{fig.hstmosaic1}. The north-facing jet is not visible in the 2019 December 8 UTC data, however, it is visible in F350LP data taken on 2020 December 9 UTC and has an angle of $\sim$60$^{\circ}$ possible due to the greater sensitivity of the 2020 December 9 F350LP observations \citep[see Fig.~6][]{Manzini2020}. 

A 1544 s F689M + F845M  composite stack taken on 2020 January 27 UTC when the comet was passing through the orbital plane is shown in the right column of Fig.~\ref{fig.hstmosaic1} indicating that the tail of 2I has a wide fan-like projection in excess of $\sim$4\arcsec~and the presence of a jet with a position angle of $\sim$250$^\circ$ and $\sim$20$^\circ$ wide while tail at an Earth and target orbital plane angle of $<$1$^\circ$. However, measuring the perpendicular velocity dispersion of dust consisting of the width of the tail is beyond the scope of this work due to the uncertainty in the timing of the ejection of the material. Observing the comet at this small orbital plane angle reveals the perpendicular to orbital vector extent of the tail perpendicular to the comet's orbital plane with minimum projection effects. The north-facing, 0$^{\circ}$ feature seen on 2019 October 12 has completely disappeared in the 2020 January 27 UTC image suggesting that projection or varying illumination effects may be concealing it or changing the level of activity during the 2020 January 27 UTC observations. The possibility of varying solar illumination on various latitudes of the comet's surface on the apparent disappearance of the northward jet is also discussed below. We do not think that the small long-term change in jet position angle between 2019 October 12 UTC and 2019 December 23 UTC is indicative of a new region becoming active, but we assume is instead due to the small changes in viewing geometry between these observations as explained in Section~\ref{s.pole}.

\subsection{Pole orientation}
\label{s.pole}

We make several assumptions to constrain the pole orientation of 2I. The first is that the rotation period of 2I is shorter than the $\sim$100 h. The longest known comet rotation period is for 1P/Halley of $\sim$7.4 d \citep[][]{Schleicher2015}, longer than the 2019 December 23-25 UTC observations, but the vast majority of known comet rotation periods are $<$24 h \citep[][]{Samarasinha2004,Kokotanekova2017}. Additional evidence is that some comets have been observed to spin up as they enter the inner Solar System due to outgassing torques caused by the increasing sublimation of volatiles producing jets exerting recoil force on the comet decreasing their spin period below this 100 h limit \citep[e.g.,][]{Samarasinha1996,Jewitt1997,Belton2011,Samarasinha2013,Bodewits2018}.

Our second assumption is that the comet is in a principal axis rotation state. One constraint on whether or not 2I is in a non-principal axis rotation state would be to observe the comet at the same viewing geometry spread out in time and observe if the jets' profile repeats themselves \citep[][]{Samarasinha2002}. However, observations of 2I at repeating viewing geometries are not available given the comet's one-time trajectory through our Solar System.

Our third assumption is that the primary jet (i.e., the south-facing jet seen in the 2019 October-December and 2020 January images), is located on or in close proximity of the spin pole and is emitting dust in a cone centered on the rotation axis. Configurations where jets emanating from the spin pole axis is common in Solar System comets \citep[eg., 17P/Borrelly][]{Farnham2002}. As seen in the observations taken on 2020 January 27 UTC during the Earth's crossing of the comet's orbital plane, the $\sim$30$^\circ$ wide jet may lie within $\sim$10$^\circ$ of the pole without affecting our pole orientation estimate. However, we caution this assumption with the fact that in the case of km-scale observations of the activity of 67P/Churyumov-Gerasimenko
 by \textit{Rossetta} showed that the majority of dust was being ejected as a jet from the northern polar Hapi region \citep[][]{Sierks2015, Schmitt2017}, however, this polar jet feature was not seen in extensive imaging observations of 67P by \textit{HST}  and ground-based assets \cite[][]{Snodgrass2017aaa}.

Our fourth assumption is a that the active region where the jet is located remains active throughout its entire rotation period. The jet could have a stationary appearance even if it is located near the equator of the comet if the activity ceases sporadically or periodically over the comet's rotation. In combination with the first two assumptions above, we expect that the comet's activity is always present through our observations.

\begin{figure}
\centering
\includegraphics[scale=0.46]{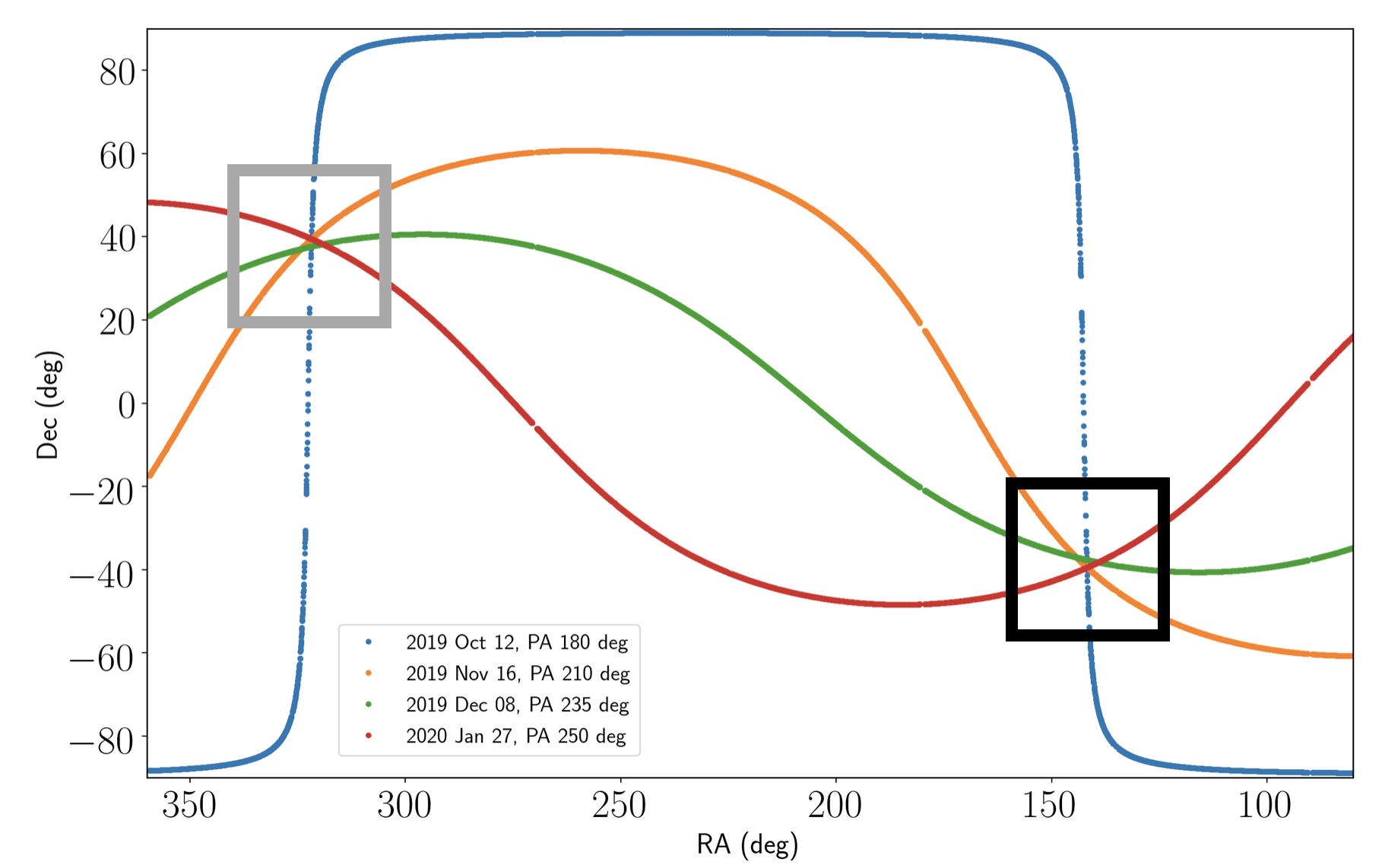}
\caption{Pole/line-of-sight planes from the 4 observing dates in Table~\ref{t:hstobs} projected on to the celestial sphere. The intersection zone of around $\alpha$ = 322$^\circ$, $\delta$ = 37$^\circ$ ($\lambda$ = 341$^\circ$, $\beta$ = 48$^\circ$) located in the grey square defines the rotation axis with an uncertainty of $\sim$10$^\circ$. The intersection zone of around $\alpha$ = 141$^\circ$, $\delta$ = -39$^\circ$ ($\lambda$ = 162$^\circ$, $\beta$ = -50$^\circ$) located in the black square defines the antipodal solution with an uncertainty of $\sim$10$^\circ$.}
\label{fig.pole}
\end{figure}

With our assumptions about the location of the jet near the spin pole of 2I, its exact position projected into three dimensions can can lie anywhere on a plane defined by the line-of-sight and jet position angle vectors for a given single observation \citep[][]{Farnham2002}. By observing 2I at different viewing geometries, the pole/line-of-sight plane will shift where the intersection of these different planes will reveal which region in inertial space that the spin pole can be located compatible with all pole/line line-of-sight planes. We have thus computed the pole/line-of-sight planes given the measured jet position angle and viewing geometry for all 6 observing dates in Table~\ref{t:hstobs} and have projected them into the celestial sphere as seen in Fig.~\ref{fig.pole}. One curve is plotted per observing date listed in for a total of six curves that intersect within $\sim$10$^\circ$ of $\alpha$ = 322$^\circ$, $\delta$ = 37$^\circ$ ($\lambda$ = 341$^\circ$, $\beta$ = 48$^\circ$ in ecliptic coordinates) with an overall uncertainty of $\sim$10$^\circ$. The close intersection of the pole/line-of-sight planes from observations at widely-spaced epochs suggest that the comet is in a mostly  principle axis rotation state \citep[][]{Farnham2002}. The uncertainty on the measured zone of convergence of the pole/line-of-sight planes is likely due to the $\sim$10$^\circ$ uncertainty on our position angle measurements and the overall scatter around the convergence point. Non-principal axis rotation could cause a scatter in the observed pole/line-of-sight planes, but the effect may be small enough to produce an effect equivalent in magnitude to the uncertainty on the convergence zone caused by our position angle measurements. There is an ambiguity caused by the projection of the jet from the spin axis pole onto the celestial sphere that prevents us from knowing the sense of rotation.

\begin{figure}
\centering
\includegraphics[scale=0.41]{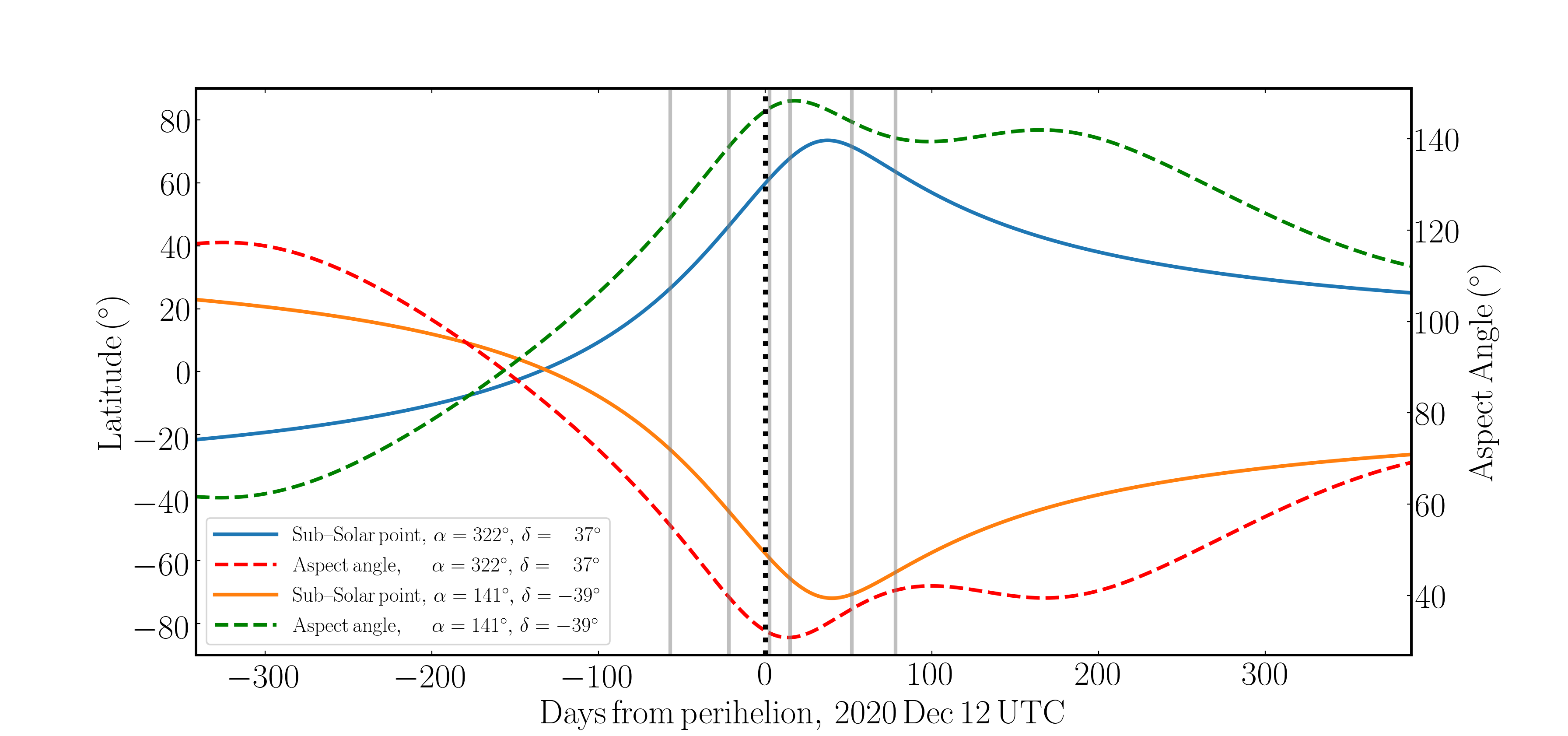}
\caption{Time since perihelion (2019 December 8 UTC) vs. sub-Solar latitude angle for 2I with the pole solution of $\alpha$ = 322$^\circ$, $\delta$ = 37$^\circ$ ($\lambda$ = 341$^\circ$, $\beta$ = 48$^\circ$) as well as for the antipodal solution of $\alpha$ = 141$^\circ$, $\delta$ = -39$^\circ$ ($\lambda$ = 162$^\circ$, $\beta$ = -50$^\circ$)  between 2019 January 01 and 2020 December 30. Also plotted is the aspect angle when viewed from the Earth, defined as is the angle between the pole and geocentric vectors centered on 2I. The time of the observations in Table~\ref{t:hstobs} are plotted as vertical grey lines and the time of perihelion is plotted as a vertical black dotted line.}
\label{fig.subsolar}
\end{figure}

The sub-Solar latitude as a function of time for our pole direction of 322$^\circ$, $\delta$ = 37$^\circ$ ($\lambda$ = 341$^\circ$, $\beta$ = 48$^\circ$) is presented in Fig.~\ref{fig.subsolar}.  Observed decreases in H$_2$O and dust production near perihelion \citep[][]{Xing2020,Cremonese2020,Bolin2020aa} may be explained by seasonal seasonal effects that are evident around the time of perihelion where the pole is becomes increasingly face-on with the Solar direction starting from $\sim$100 days before perihelion. This may have resulted in a decrease in temperature of relatively water-rich areas observed as a corresponding decrease in dust production and H$_2$O-driven activity. At the same time, spectroscopic observations of 2I from \textit{HST} revealed that the CO production rate was increasing as the comet was nearing perihelion \citep[][]{Bodewits2020} which may suggest that the pole is rich in CO that was becoming more active as the pole was increasingly heated by the Sun.

Also presented in Fig.~\ref{fig.subsolar} is the sub-Solar latitude for the antipodal solution of $\alpha$ = 141$^\circ$, $\delta$ = -39$^\circ$ ($\lambda$ = 162$^\circ$, $\beta$ = -50$^\circ$). The antipodal solution isn't exactly 180$^\circ$ from the main pole solution due to the presence of scatter in the intersection of the Pole/line-of-sight planes in the antipodal direction as seen in Fig.~\ref{fig.pole}. As the pole containing the primary jet becomes more face on with perihelion, its opposite pole becomes corresponding less face on entering polar night just after perihelion. As noted here and  in \textit{HST} images taken between 2019 October 12 UTC and 2019 December 6 UTC \citep[see Fig.~6][]{Manzini2020} there is a second jet opposite the primary jet with a position angle $\sim$180$^\circ$ in difference. Projection effects may be distorting the appearance of this second jet, but its $\sim$180$^\circ$ difference in position angle from the primary jet may suggest it is located near the opposite hemisphere. As described by \citet[][]{Manzini2020} and \citet[][]{Kim2020}, this second jet appears to decrease in intensity in \textit{HST} images as the comet approaches perihelion completely disappearing in images taken on 2020 January 3 UTC. This may be explained by a decrease in activity on the pole opposite the primary jet as it enters polar night.

The pole solution of \citet[][]{Manzini2020}, $\alpha$ = 260$^\circ$, $\delta$ = -35$^\circ$ ($\lambda$ = 260$^\circ$, $\beta$ = -11$^\circ$) differs significantly from our result. To constrain the pole of 2I, \citet[][]{Manzini2020} interpret the $\sim$180$^\circ$ difference between the two jets seen in the 2019 October 12 UTC observations as the primary jet and secondary jets being located within located within $\sim$10$^\circ$ of the comet's equator and that the spin axis is lying within the plane of the sky with a position angle of $\sim$100$^\circ$/280$^\circ$. However, the near-equatorial position of the jets assumed by \citet[][]{Manzini2020} are not compatible with our assumptions about the comet's relatively short rotation period due to their stationary appearance in \textit{HST} images discussed below in Section~\ref{s.conclusions} and due to the polar appearance of jets of some comets. However, we do note that long-term observations of the activity of 67P by the \textit{Rosetta} mission did not reveal jets occurring simultaneously from opposite poles \citep[][]{Sierks2015}.

In addition, \citet[][]{Kim2020} calculates a pole solution of $\alpha$ = 205$^\circ$, $\delta$ = 52$^\circ$ ($\lambda$ = 172$^\circ$, $\beta$ = 56$^\circ$) differing significantly from our result. They interpret the anisotropic morphology of the coma as consistent with the location of the peak mass loss located on the surface of the comet displaced from its local noon due to thermal lag \citep[e.g.,][]{Samarasinha2004}. Therefore the jet's location in the thermal lag scenario is defined by its proximity towards the sub-Solar point and can therefore be located at a location off the spin pole axis unless pole is pointed in the direction of the Sun during its passage through the Solar System. According to Fig. 7 from \citep[][]{Kim2020}, the sub-Solar point defined by their pole solution is located within $\sim$20$^\circ$ of the comet's equator during the comets travel through the Solar System covered by the \text{HST} which differs from our result. In addition, we note that the different assumptions in interpreting the pole solution in \citet[][]{Manzini2020}, \citet[][]{Kim2020} and in this work lead to different pole solutions which may suggest current \textit{HST} observations do not constrain a unique pole solution.

\subsection{Photometry and lightcurves}
\label{sec:photometry}
We perform an independent analysis and calibration of the  2019 October 12 data presented in \citet[][]{Jewitt2019hst}. We calculate the photometry on according to the aperture correction and zero-points determined for the F350LP filter \citep[][]{Deustua2017} and measure the photometry of 2I in the \textit{HST} images using a circular aperture with an equivalent 10,000 km aperture resulting in $m_{\mathrm{F350LP}}$ = 17.43 $\pm$ 0.01. Ground-based $V$ filter observations of 2I by the MLO 1.0-m were conducted near simultaneously to the \textit{HST} observations on 2019 October 12 UTC with the comet's $V$ = 17.55 $\pm$ 0.04 \citep[][]{Bolin2020aa} which we used to estimate the color between the F350LP and the $V$ filters, F350LP $-$ $V$ = $-$0.12 $\pm$ 0.04.

We then use the F350LP - V band color to determine the V magnitude in the 2019 October 12 observations using an aperture radius of 0.2\arcsec radius and a sky and outer coma subtraction annulus between 0.2-0.8$\arcsec$ resulting in a V magnitude of 21.51 $\pm$ 0.04. The smaller 0.2\arcsec~radius aperture enabled by \textit{HST} allows for the removal of orders of magnitude more light from the coma within the vicinity of the nucleus compared to ground-based observations enhancing the contrast between the coma nucleus \citep[e.g.][]{Jewitt2019b}. We similarly use the $V$-$r$ = 0.45, $V$- $i$ = 0.62  and $V$-$z$ = 0.42 colors of 2I/Boriosv taken from ground-based facilities \citep[][]{Bolin2020aa} and absolute magnitude of the Sun from \citet[][]{Willmer2018} to calculate the equivalent V magnitude of 2I from the F689M and F845M data.

Using the following equation and the V magnitudes calculated above
\begin{equation}
\label{eqn.brightness}
H = V - 5\, \mathrm{log_{10}}(r_h \Delta) - \Phi(\alpha)
\end{equation}
we calculate the absolute magnitude H according to the heliocentric distance $r_h$ in au and geocentric distance $\Delta$ in au from Table ~\ref{t:hstobs} and $\Phi(\alpha)$ = 0.04$\alpha$, where $\alpha$ is the phase angle in degrees and 0.04 is the phase coefficient in magnitudes/degree, resulting in H magnitude of 16.59 $\pm$ 0.03 for the 2019 October 12 observations date. The rest of the H magnitude measurements for the data sets taken on other dates are listed in Table ~\ref{t:hstobs}.  We note that the errors on the H magnitudes may be underestimated in part due to the unknown phase function of 2I which might have a slightly larger phase coefficient than 0.04 magnitudes/degree.

As discussed in \citet[][]{Jewitt1991}, the dilution of light from the nucleus by dust in the coma can dampen the variability in a comet lightcurve for timescales shorter than the crossing time of dust within the scale of the photometric aperture. For 2I, due to the density and slow crossing time of dust within its coma, measuring any short-term lightcurve variations on the order of hours to 10's of hours caused by the rotation of the comet's nucleus is difficult at the coarse resolution of ground-based observations \citep[][]{Jewitt2019}. We thus search instead for short-term variations in the lightcurve such as due to the rotation of the nucleus using the high-resolution WFC3/UVIS taken on 2019 October 12 UTC and in the data taken on 2019 December 23-25 UTC. At distances close to the surface of the comet, the speed of small, efficiently light scattering dust particles coupled directly to outflowing gas is approximately the speed of sound in gas \citep[][]{Jewitt1991}, 0.43 km/s at the black body temperature of the gas, 181 K, at the heliocentric distance of the comet of 2.37 au on 2019 October 12 UTC and 197 K at the heliocentric distance of the comet of 2.03 au on 2019 December 24 UTC. For ground-based observations taken with a 10,000 km aperture, this translates into a dust crossing time of$\gtrsim$10 h. We can use the superb 0.04\arcsec/pixel (or $\sim$60 km) resolution of the WFC3/UVIS data to measure the brightness of the comet with a smaller aperture enabling shorter coma dust dampening timescales.

For this analysis, we use a 0.2\arcsec~radius aperture centered on the peak of the comet's brightness profile with a contiguous sky and outer coma subtraction aperture 0.2\arcsec-0.8\arcsec. The equivalent distance spanning 0.2\arcsec~at a distance of 2.79 au, the geocentric distance of 2I on 2019 October 12 UTC, is $\sim$400 km and $\sim$300 km at the geocentric distance of the comet of 1.93 au on 2019 December 24, 2019 UTC. These equivalent distances in 2019 October and November correspond to a crossing time of $>$10 h assuming a 1-10 m/s dust ejection velocity for mm-sized dust particles, which dominate the 2I dust cross-section as evidenced by ground and space-based observations \citep[][]{Cremonese2020,Hui2020a,Kim2020}. While the smaller 0.2\arcsec~radius aperture enhances the contrast between the region containing the nucleus of the comet and the rest of the coma increasing the potential for measuring the variability of the lightcurve from the rotation of the comet's nucleus \citep[e.g.,][]{Lamy1998a,Lamy1998b}, the slow crossing time of the dust may prevent the measurement of periodic lightcurve variations shorter than $\sim$10 h.

\begin{figure}
\centering
\includegraphics[scale=0.28]{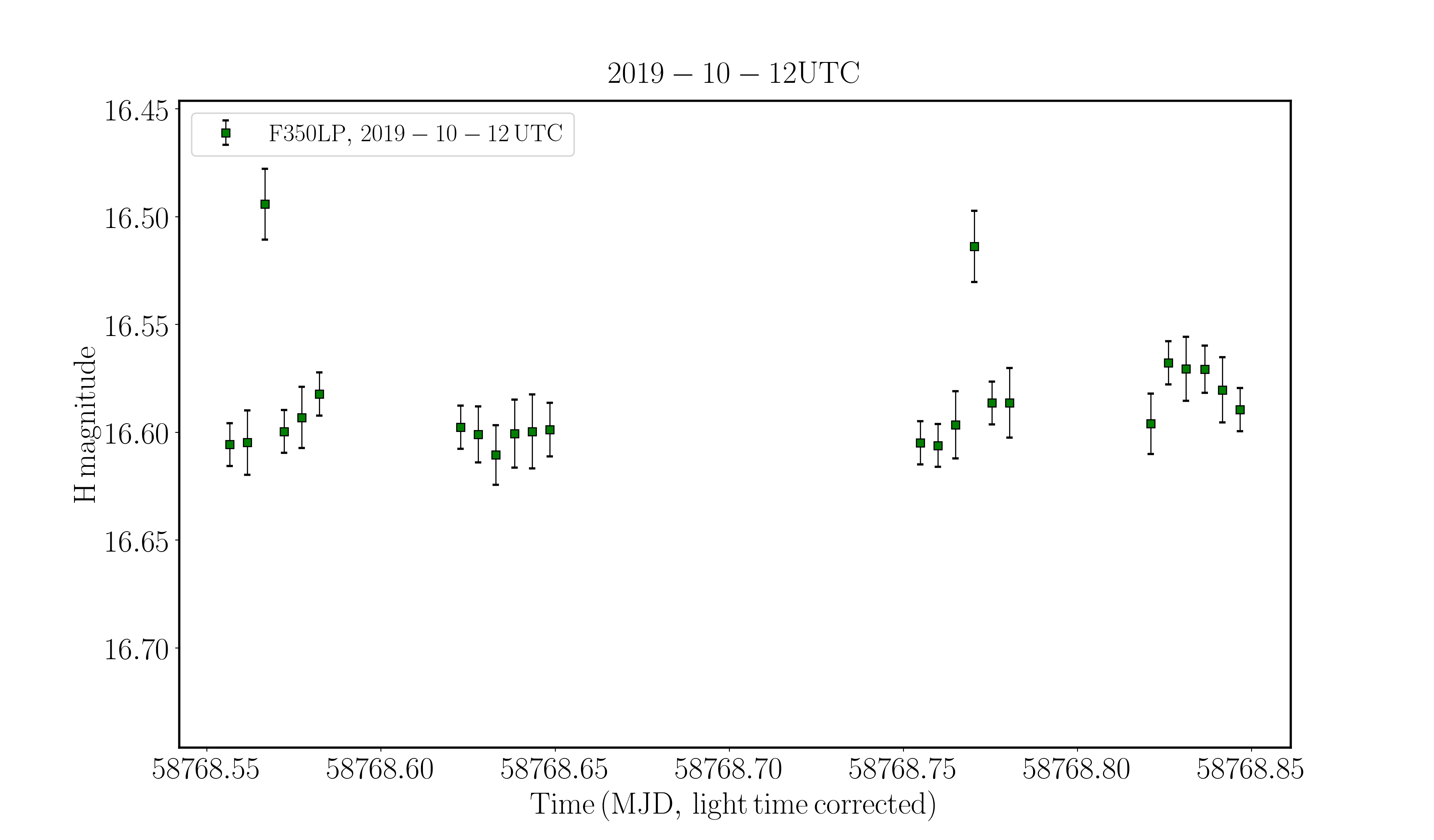}
\includegraphics[scale=0.28]{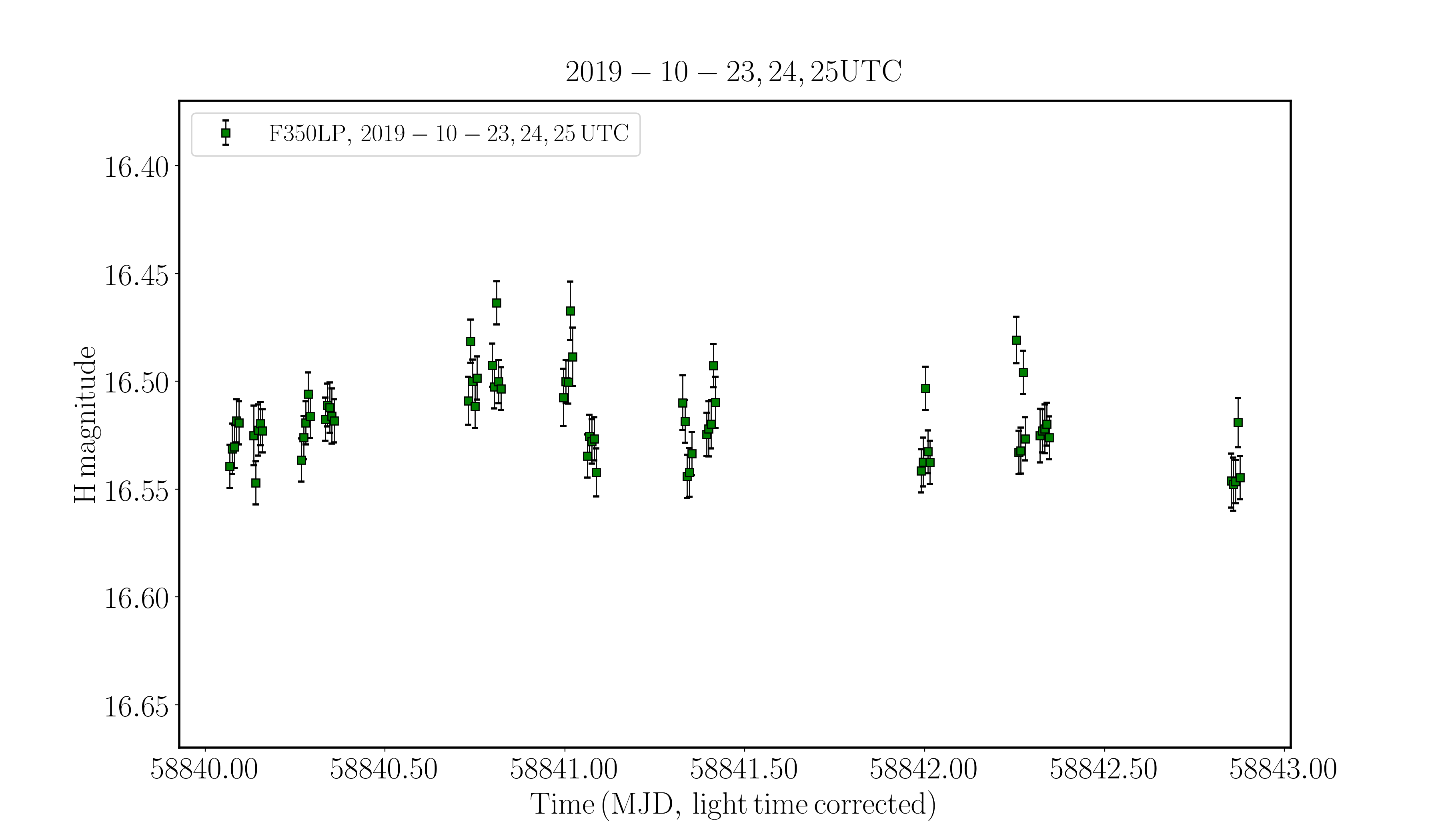}
\caption{Top and bottom panels: WFC3/UVIS F350LP lightcurves of 2I taken on 2019 October 12 UTC and 2019 December 23-25 UTC. The photometry was measured using a 0.2\arcsec~radius aperture radii. The overestimated data points in the first set of six and third set of six measurements are due to background stars coming into contact with the aperture centered on the brightest point of the comet. The error bars on the data points are equal to their 1 $\sigma$ photometric uncertainties. The data have been detrended and points affected by trailed background stars have been removed.}
\label{fFiig:hstlightcurve}
\end{figure}

Our \textit{HST} photometric measurements of 2I from \textit{HST} observations taken on 2019 October 12 UTC are presented in Table~\ref{t.photometry1} and in the top panel of Fig.~\ref{fFiig:hstlightcurve}. Our results are consistent with those of \citet[][]{Jewitt2019hst} showing only small variations in the lightcurve comparable to the $\gtrsim$0.01-0.02 mag photometric uncertainties of individual datapoints which may be due to the slow crossing time of dust discussed above. The time of the observations has been corrected for light-travel time and the photometry has been kept in F350LP magnitude.

As an additional check, we have investigated the photometry of an additional set of observations of 2I were obtained with \textit{HST} \citep[HST GO 16043, PI][]{MeechHST2019b} that consisted of 14 orbits spread out over a $\sim$70 h period between 2019 December 23 01:54:45 UTC and 2019 December 25 21:24:46 UTC. Observations during each of the 14 orbits on 2019 December 23-25 UTC resulted in 70 separate images with an equivalent exposure time of 380 s in the F350LP filter presented in. The lightcurve consisting of data points taken during 2019 December 23-25 UTC observations are presented in Table~\ref{t.photometry2} and in the bottom panel of Fig.~\ref{fFiig:hstlightcurve}. There appears to be some variability on the scale of $\sim$0.05 magnitude, at SNR 2.5 to 5.0.

\begin{figure}
\centering
\includegraphics[scale=0.24]{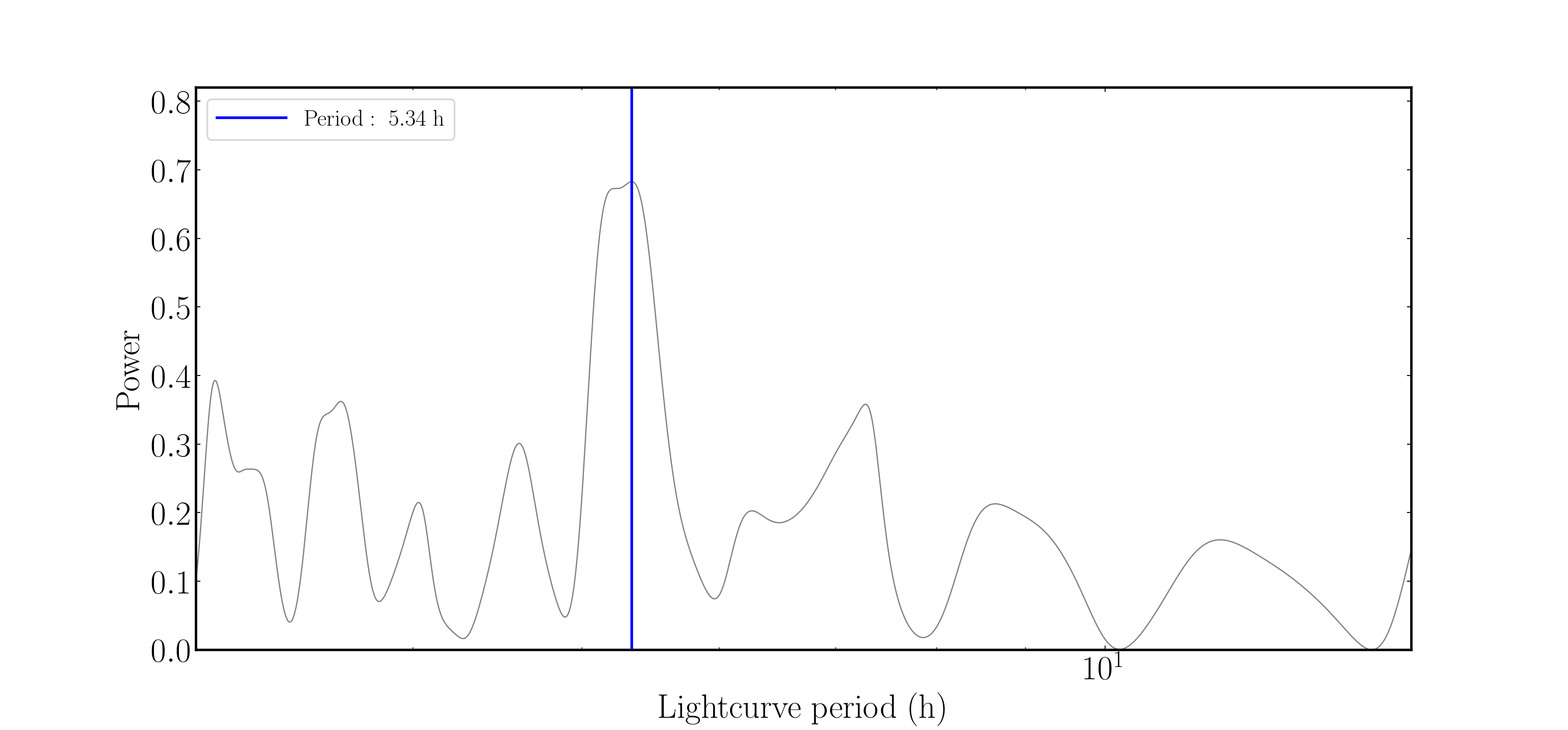}
\includegraphics[scale=0.24]{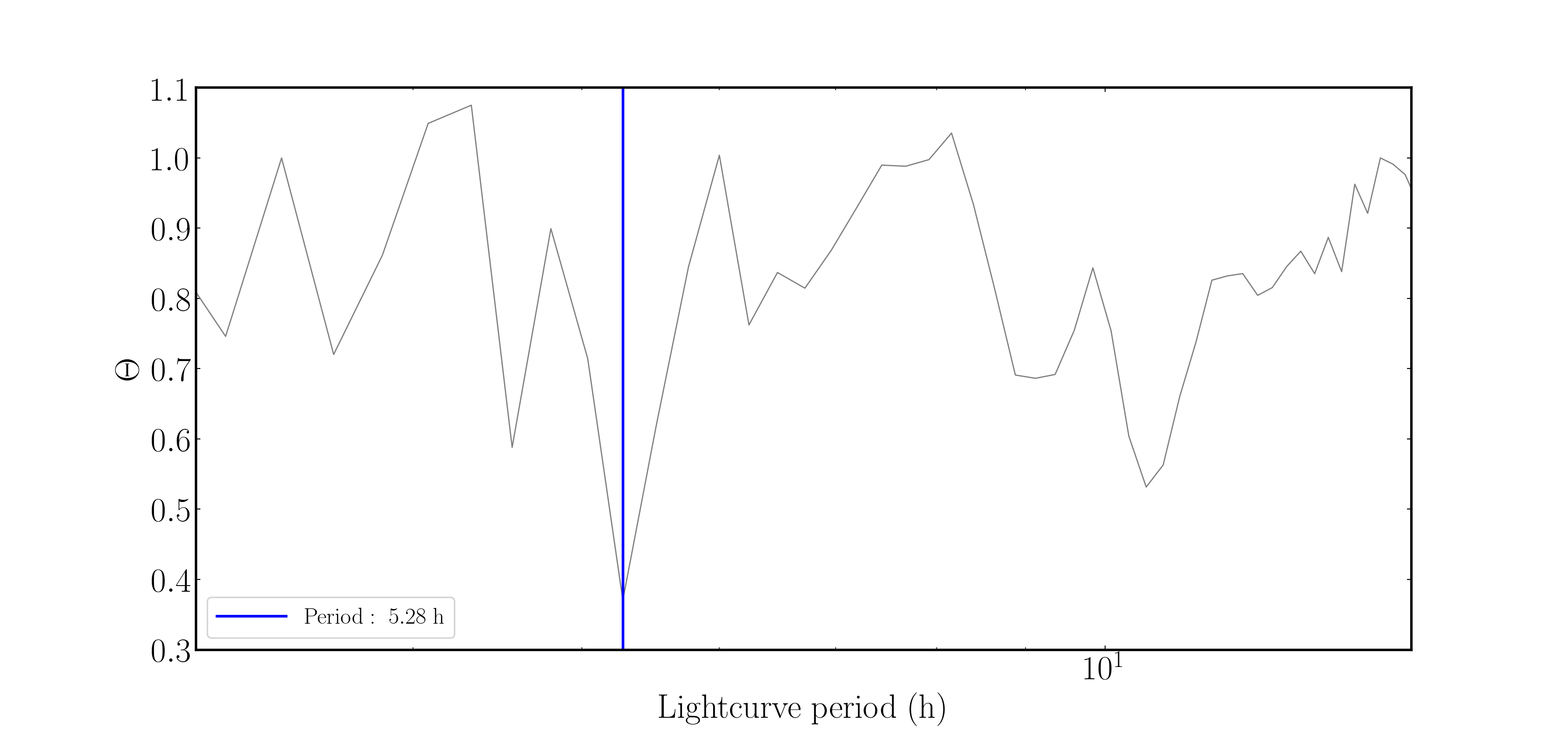}
\includegraphics[scale=0.24]{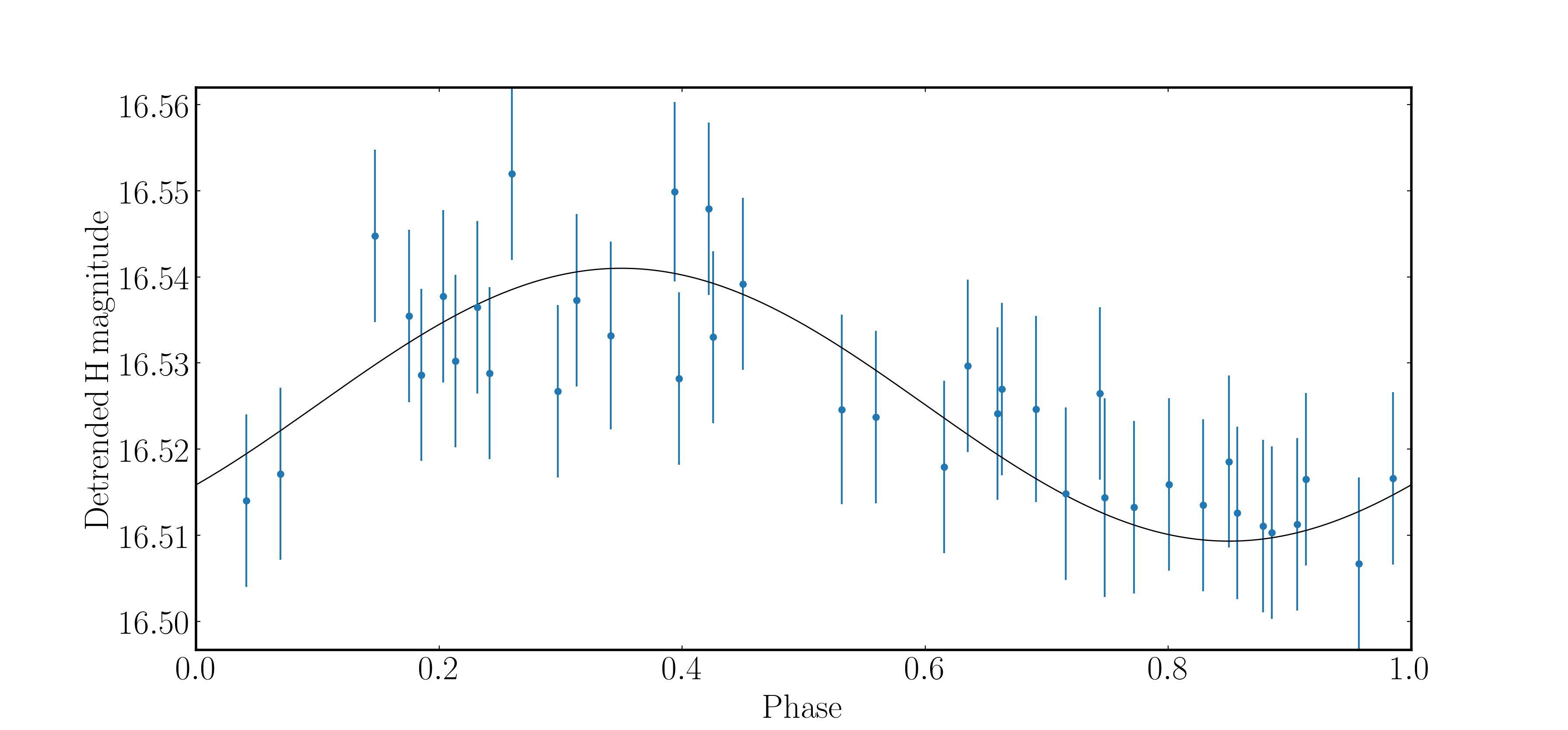}
\includegraphics[scale=0.24]{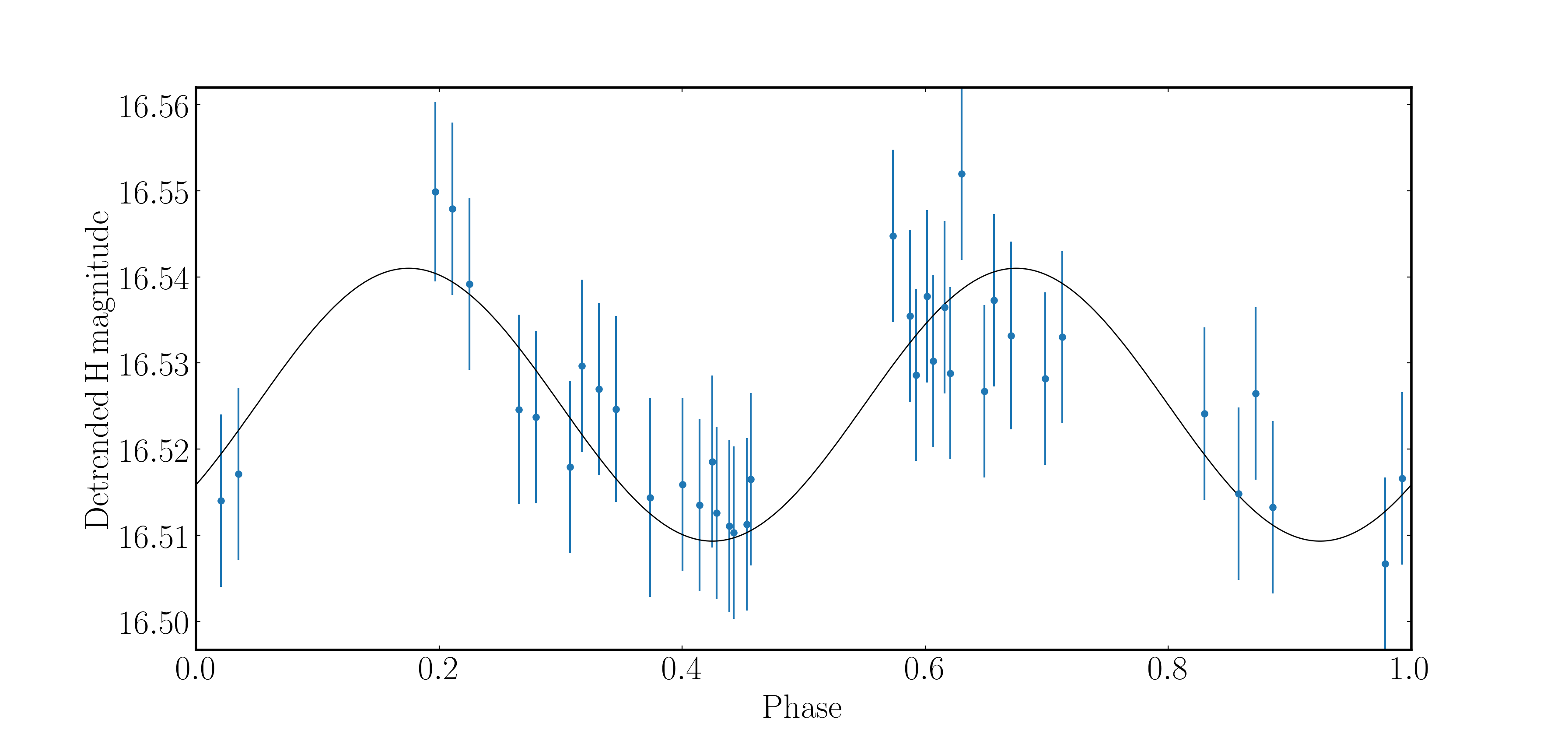}
\caption{Top panel: Lomb-Scargle periodogram of lightcurve period vs spectral power \citep[][]{Lomb1976} for the WFC3/UVIS F350LP lightcurve data from the 2019 December 23-25 UTC observations. A peak in the power is located at double-peaked rotation period of 10.67 h with a FWHM of $\sim$0.1 h. Second panel: Phase dispersion minimization analysis of lightcurve rotation period vs. $\Theta$ metric \citep[][]{Stellingwerf1978}. The $\Theta$ metric is minimized at double-peaked rotation periods of 10.6 h consistent with the 10.67 h rotation period fond with the Lomb-Scargle Periodogram. Third and fourth panel: phased WFC3/UVIS F350LP lightcurve using data from the 2019 December 23-25 UTC observations corresponding to a single-peak lightcurve period of 5.34 h and a double-peak lightcurve period of 10.68 h. }
\label{fFiig:hstperiod}
\end{figure}

Removing the linear trend fit over the data taken during the 2019 December 23-25 UTC and removing data points affected by background stars, we apply the Lomb-Scargle periodogram \citep[][]{Lomb1976} to the data which is displayed in the top panel of Fig.~\ref{fFiig:hstperiod}. Removal of the linear trend over the three days observing period will affect the determination of rotation periods that are on multi-day time scales, but do not  periodicities shorter than $\sim$10 h time scales. The highest peak in the lightcurve period/rotation period vs. spectral power curve is located at $\sim$5.3 h with a formal significance of $p \simeq$ 10$^{-6}$. We apply bootstrap estimation \citep[][]{Press1986} of the uncertainties by removing $\sqrt{N}$ data points from the time series lightcurve and repeating our periodogram estimation of the rotation period 10,000 times resulting in a 1~$\sigma$ uncertainty estimate of $\sim$0.1 h. As an independent check of our results obtained with the Lomb-Scargle periodogram, we apply phase dispersion minimization analysis to our data \citep[][]{Stellingwerf1978} and obtain a result of $\sim$10.6 h compatible with the rotation period estimate obtained with the Lomb-Scargle periodogram as seen in the second panel of Fig.~\ref{fFiig:hstperiod}.

We caution that although our periodogram analysis shows evidence for a periodic signal, it may be suspect because the single-peaked and double-peaked phased data as seen in the third panel indicate a small amplitude of only $\sim$0.05 magnitudes (third and bottom panels, Fig.~\ref{fFiig:hstperiod}), comparable to within a factor of a few of the errors on the individual data points. Even though small lightcurve amplitude might be expected in this case where the nucleus was not detected in the coma of this object \citep[][]{Bolin2020aa, Jewitt2019hst,Kim2020} due to the source within our photometric aperture being dominated by the coma's dust \citep[e.g.,][]{Hsieh2012}, we
 note that the small lightcurve amplitude seen in observations of 2I is in stark contrast with the case of 1I/`Oumuamua, which was observed to have $>$2 magnitude variations in its $\sim$8 h rotation period lightcurve \citep[][]{Knight2017,Bolin2018}. In addition, it has been shown in the case of 67P/Churyumov-Gerasimenko that significant variations of cometary activity can occur on hour timescales comparable crossing time of dust within the 0.2\arcsec~radius aperture used to measure the photometry of the lightcurve \citep[][]{Lin2017}.

\section{Discussion and Conclusions}
\label{s.conclusions}

Although many observations of 2I have already occurred \citep[e.g.,][]{Jewitt2019, Fitzsimmons2019, Bolin2020aa,Hui2020a}, our understanding of 2I and its context within the greater interstellar comet populations is only beginning to unfold. With the best-available spatial data of 2I, we have determined the existence of jet features in the coma surrounding the nucleus between 2019 October 12 UTC to 2020 January 27 UTC before and after its perihelion passage on 2019 December 8 UTC. Close time-series data taken on 2019 December 23-25 UTC show a possible $\sim$0.05 magnitude lightcurve variations, however it is likely that these variations are not due to the rotation of the nucleus as a result of the comet's cross-section being dominated by dust. Therefore, it may not be possible to constrain the rotation period of 2I with current \textit{HST} observations.

\begin{figure}
\centering
\includegraphics[scale=0.41]{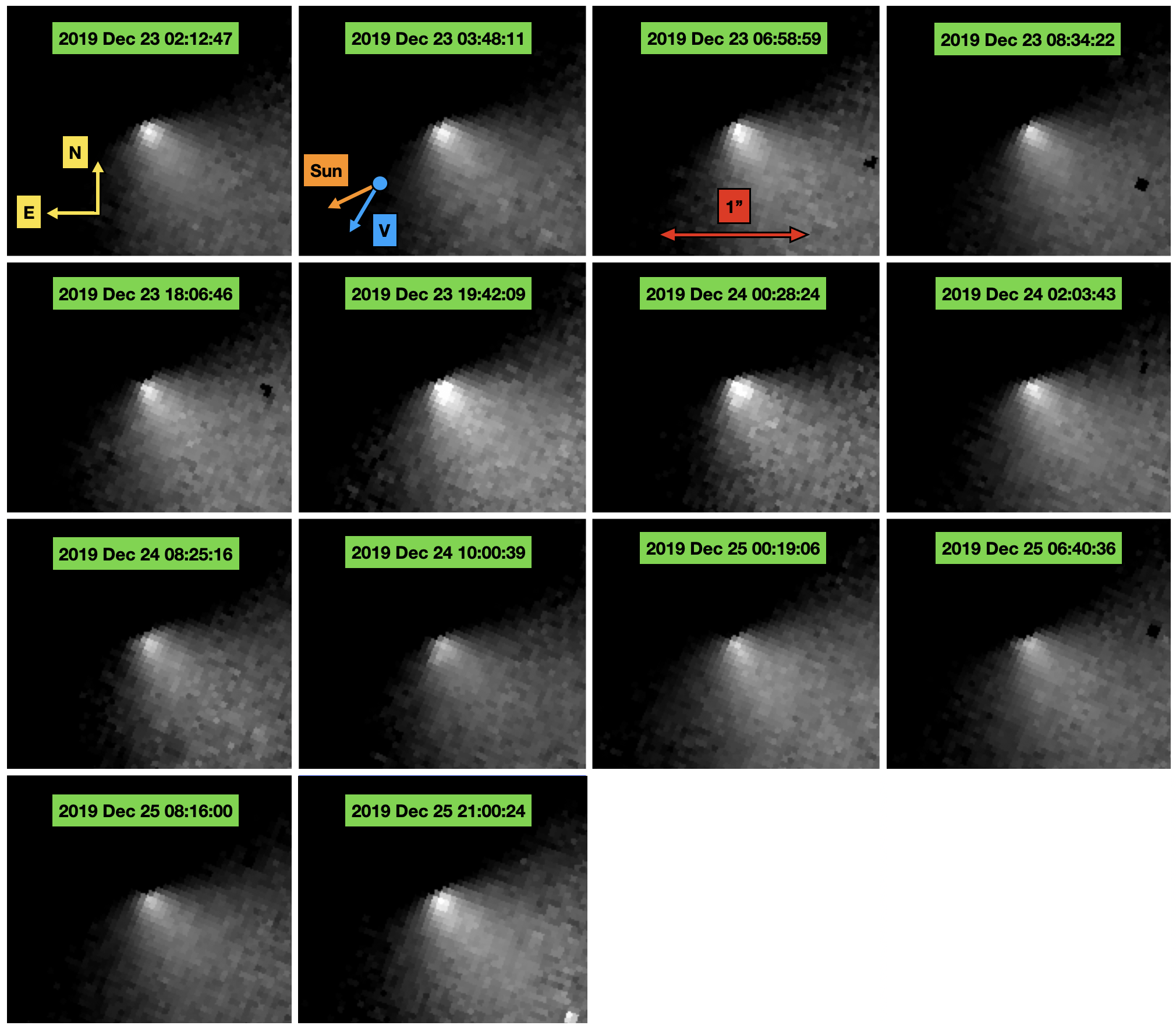}
\caption{Mosaic of F350LP WFC3/UVIS images obtained over 14 orbits between 2019 December 23 02:12:47 UTC and 2019 December 25 21:00:24 UTC \citep[HST GO 16043, ][]{MeechHST2019b}. Each panel is a median stack of five 380 s exposures obtained during each of the orbits resulting in an equivalent exposure time of 1900 s. The radial profile of the coma has been removed from the detection of 2I in each median stack. The cardinal direction vectors, the solar and orbital velocity and spatial scale are indicated.}
\label{fig.hstmosaic3}
\end{figure}

We have used the $\sim$5 month time span to determine the spin pole position of 2I finding that the data are consistent with the pole direction being towards $\alpha$ = 322$^\circ$, $\delta$ = 37$^\circ$ ($\lambda$ = 341$^\circ$, $\beta$ = 48$^\circ$) assuming that the jet is located near the comet's rotation pole. Our assumptions are supported by checking the appearance of the jet in the 2019 December 23-25 UTC data seen in Fig.~\ref{fig.hstmosaic3} which does not appear to significantly change its position angle over the $\sim$70 h time span of the observations. The tight convergence of our pole/line-of-sight planes presented in Fig.~\ref{fig.pole} may imply that the comet is in a simple rotation state \citep[][]{Farnham2002} unlike 1I/`Oumuamua which was observed to be in a tumbling rotational state \citep[][]{Fraser2018}. However, given that three different assumptions lead to three different pole solutions as discussed in Section~\ref{s.pole}, we conclude that current \textit{HST} observations of 2I may not provide a unique pole solution until additional observations are made that may provide additional evidence favoring particular assumptions.

If we assume that it is a rotating body whose shape is approximated by a triaxial prolate shape with dimensions with dimensions, $a$:$b$:$c$ where $b$ $\geq$ $a$ $\geq$ $c$ similar to Solar System comets \citep[e.g.,][]{Binzel1989,Samarasinha2004}, a small lightcurve amplitude of $\sim$0.05 may imply a low $b/a$ axial ratio of $\sim$1 where $b/a \; = \; 10^{0.4 A}$ where $A$ is the peak-to-trough lightcurve amplitude \citep[][]{Binzel1989}.  The aspect angle between the observer's line of sight and the rotational pole can affect the observed lightcurve amplitude where an object viewed pole on will appear to have a small lightcurve amplitude as the light-reflective cross-section when viewed at this angle is unchanging regardless of its shape \citep[][]{Barucci1982}. Although the light scattering cross-section of the nucleus is dust dominated in our observations even at the resolution enabled by \textit{HST} and the lightcurve amplitude is very small at $\sim$0.05, we attempt to estimate a lower limit to the $b/a$ taking into account the aspect angle that 2I was viewed at during the 2019 December 23-25 UTC observations using the following
\begin{equation}
\label{eq.viewingmag}
Ê \Delta m_{\rm diff} = 1.25\,\mathrm{log}\left( \frac{b^2\cos^2\theta \; + \; c^2\sin^2\theta}{a^2\cos^2\theta \; + \; c^2\sin^2\theta} \right )
\end{equation}
where $a$, $b$ and $c$ are the dimensions of 2I and $\theta$ is the aspect angle in degrees \citep[][]{Thirouin2016}. We will assume 1$\lesssim$$b/a$$\lesssim$2 typical for Solar System comets \citep[][]{Lamy2004} and that $a$ = $c$ for a prolate triaxial elipsoid. On 2019 December 23-25 UTC, 2I was observed at an aspect angle $\theta$ = 30$^{\circ}$ as seen in Fig.~\ref{fig.subsolar} which combined with the above assumption results on 2I's triaxial shape results in a lightcurve amplitude correction of $\sim$0.4 and $b/a \; = \; 10^{0.4A}$ $\gtrsim$ 1.5. 

However, a nearly spherical body with a homogeneously reflective surface viewed equatorially would have a small lightcurve amplitude \citep[][]{Harris2014, Hanus2018}. In either of these cases, due to its dust-dominated coma observed in current \textit{HST} data, it is not possible to tell if  2I has an extremely elongated shape like 1I/`Oumuamua \citep[][]{Bolin2018,Seligman2020}. Therefore, we present our results as demonstrating the non-uniqueness of inferred rotation properties of 2I using \textit{HST} observations in its pre-Solar System encounter, pre-outburst rotational state \citep[][]{Jewitt2020aadfasd}. Subsequent observations of 2I will be required to place more stringent constraints on its rotational state though may only be applicable to its post-outburst state and if the activity evolves so that the light scattering cross-section is less dust-dominated. Therefore, it may not be possible to fully constrain the rotation of 2I in its pre-Solar System encounter state.

\section*{Acknowledgements}
Based on observations with the NASA/ESA Hubble Space Telescope obtained from the Data Archive at the Space Telescope Science Institute, which is operated by the Association of Universities for Research in Astronomy, Incorporated, under NASA contract NAS5-26555. Support for Program number (GO 16040) was provided through a grant from the STScI under NASA contract NAS5-26555.

We would like to acknowledge the anonymous reviewer whose helpful comments significantly improved the quality of this manuscript.

While this manuscript was in review, two other studies by \citet[][]{Manzini2020} and \citet[][]{Kim2020} focusing on the observations of 2I by \textit{HST} and its pole solution were accepted and or submitted on arxiv respectively. The results in both papers complement our work and enhance the scientific discussion in this manuscript.

We would like to thank R. Jedicke and G. Helou for helpful discussion on the interpretation of low amplitude lightcurve data. We would also like to thank Y. R. Fernandez for helpful discussion on the interpretation of the comet coma data and for useful comments on constraining the pole. We thank D. Bodewits for advice in planning the \textit{HST} observations and data reduction.

\section*{Data Availability}

The data underlying this article will be shared on reasonable request to the corresponding author. The data are also available at the Mikulski Archive for Space Telescopes (MAST).

\bibliographystyle{mnras}
\bibliography{ms}

\begin{longtable}{|c|c|c|c|}
\caption{Summary of comet 2I photometry taken on 2019 October 12 UTC.\label{t.photometry1}}\\
\hline
Date$^1$ & Filter$^2$ & Exp$^3$&$H^4$ \\
UTC&&(s)&\\
\hline
\endfirsthead
\multicolumn{4}{c}%
{\tablename\ \thetable\ -- \textit{Continued from previous page}} \\
\hline
Date$^1$ & Filter$^2$ & Exp$^3$&$H^4$ \\
UTC&&(s)&\\
\hline
\endhead
\hline \multicolumn{4}{r}{\textit{Continued on next page}} \\
\endfoot
\hline
\endlastfoot
58768.5565711 & F350LP & 260 s & 16.61 $\pm$ 0.02 \\
58768.5616289 & F350LP & 260 s & 16.60 $\pm$ 0.01 \\
58768.5666868 & F350LP & 260 s & 16.49 $\pm$ 0.02 \\
58768.5721266 & F350LP & 260 s & 16.60 $\pm$ 0.02 \\
58768.5771845 & F350LP & 260 s & 16.59 $\pm$ 0.01 \\
58768.5822424 & F350LP & 260 s & 16.58 $\pm$ 0.01 \\
58768.6227863 & F350LP & 260 s & 16.60 $\pm$ 0.01 \\
58768.6278442 & F350LP & 260 s & 16.60 $\pm$ 0.01 \\
58768.6329021 & F350LP & 260 s & 16.61 $\pm$ 0.01 \\
58768.6383419 & F350LP & 260 s & 16.60 $\pm$ 0.01 \\
58768.6433998 & F350LP & 260 s & 16.60 $\pm$ 0.01 \\
58768.6484576 & F350LP & 260 s & 16.60 $\pm$ 0.01 \\
58768.7547887 & F350LP & 260 s & 16.60 $\pm$ 0.02 \\
58768.7598465 & F350LP & 260 s & 16.61 $\pm$ 0.01 \\
58768.7649044 & F350LP & 260 s & 16.60 $\pm$ 0.02 \\
58768.7703442 & F350LP & 260 s & 16.51 $\pm$ 0.02 \\
58768.7754021 & F350LP & 260 s & 16.59 $\pm$ 0.02 \\
58768.78046 & F350LP & 260 s & 16.59 $\pm$ 0.01 \\
58768.8209924 & F350LP & 260 s & 16.60 $\pm$ 0.01 \\
58768.8260502 & F350LP & 260 s & 16.57 $\pm$ 0.01 \\
58768.8311081 & F350LP & 260 s & 16.57 $\pm$ 0.01 \\
58768.8365479 & F350LP & 260 s & 16.57 $\pm$ 0.02 \\
58768.8416058 & F350LP & 260 s & 16.58 $\pm$ 0.01 \\
58768.8466637 & F350LP & 260 s & 16.59 $\pm$ 0.01 \\
\hline
\caption{Columns: (1) observation date correct for light travel time; (2) \textit{HST}/WFC3 Filter; (3) Exposure time (4) H magnitude with 1 $\sigma$ uncertainties}
\end{longtable}

\begin{longtable}{|c|c|c|c|}
\caption{Summary of comet 2I photometry taken on 2019 December 23-25  UTC.\label{t.photometry2}}\\
\hline
Date$^1$ & Filter$^2$ & Exp$^3$&$H^4$ \\
UTC&&(s)&\\
\hline
\endfirsthead
\multicolumn{4}{c}%
{\tablename\ \thetable\ -- \textit{Continued from previous page}} \\
\hline
Date$^1$ & Filter$^2$ & Exp$^3$&$H^4$ \\
UTC&&(s)&\\
\hline
\endhead
\hline \multicolumn{4}{r}{\textit{Continued on next page}} \\
\endfoot
\hline
\endlastfoot
58840.0684946 & F350LP & 380 s & 16.54 $\pm$ 0.01 \\
58840.0747562 & F350LP & 380 s & 16.53 $\pm$ 0.01 \\
58840.0810178 & F350LP & 380 s & 16.53 $\pm$ 0.01 \\
58840.0872794 & F350LP & 380 s & 16.52 $\pm$ 0.01 \\
58840.0935409 & F350LP & 380 s & 16.52 $\pm$ 0.01 \\
58840.1347446 & F350LP & 380 s & 16.53 $\pm$ 0.01 \\
58840.1410062 & F350LP & 380 s & 16.55 $\pm$ 0.01 \\
58840.1472678 & F350LP & 380 s & 16.52 $\pm$ 0.01 \\
58840.1535294 & F350LP & 380 s & 16.52 $\pm$ 0.01 \\
58840.1597909 & F350LP & 380 s & 16.52 $\pm$ 0.01 \\
58840.2672446 & F350LP & 380 s & 16.54 $\pm$ 0.01 \\
58840.2735062 & F350LP & 380 s & 16.53 $\pm$ 0.01 \\
58840.2797678 & F350LP & 380 s & 16.52 $\pm$ 0.01 \\
58840.2860294 & F350LP & 380 s & 16.51 $\pm$ 0.01 \\
58840.2922909 & F350LP & 380 s & 16.52 $\pm$ 0.01 \\
58840.3334831 & F350LP & 380 s & 16.52 $\pm$ 0.01 \\
58840.3397446 & F350LP & 380 s & 16.51 $\pm$ 0.01 \\
58840.3460062 & F350LP & 380 s & 16.51 $\pm$ 0.01 \\
58840.3522678 & F350LP & 380 s & 16.52 $\pm$ 0.01 \\
58840.3585294 & F350LP & 380 s & 16.52 $\pm$ 0.01 \\
58840.7309831 & F350LP & 380 s & 16.51 $\pm$ 0.01 \\
58840.7372446 & F350LP & 380 s & 16.48 $\pm$ 0.01 \\
58840.7435062 & F350LP & 380 s & 16.5 $\pm$ 0.01 \\
58840.7497678 & F350LP & 380 s & 16.51 $\pm$ 0.01 \\
58840.7560294 & F350LP & 380 s & 16.5 $\pm$ 0.01 \\
58840.7972215 & F350LP & 380 s & 16.49 $\pm$ 0.01 \\
58840.8034831 & F350LP & 380 s & 16.5 $\pm$ 0.01 \\
58840.8097446 & F350LP & 380 s & 16.46 $\pm$ 0.01 \\
58840.8160062 & F350LP & 380 s & 16.5 $\pm$ 0.01 \\
58840.8222678 & F350LP & 380 s & 16.5 $\pm$ 0.01 \\
58840.9960062 & F350LP & 380 s & 16.51 $\pm$ 0.01 \\
58841.0022678 & F350LP & 380 s & 16.5 $\pm$ 0.01 \\
58841.0085294 & F350LP & 380 s & 16.5 $\pm$ 0.01 \\
58841.0147909 & F350LP & 380 s & 16.47 $\pm$ 0.01 \\
58841.0210525 & F350LP & 380 s & 16.49 $\pm$ 0.01 \\
58841.0621983 & F350LP & 380 s & 16.53 $\pm$ 0.01 \\
58841.0684599 & F350LP & 380 s & 16.53 $\pm$ 0.01 \\
58841.0747215 & F350LP & 380 s & 16.53 $\pm$ 0.01 \\
58841.0809831 & F350LP & 380 s & 16.53 $\pm$ 0.01 \\
58841.0872446 & F350LP & 380 s & 16.54 $\pm$ 0.01 \\
58841.3271636 & F350LP & 380 s & 16.51 $\pm$ 0.01 \\
58841.3334252 & F350LP & 380 s & 16.52 $\pm$ 0.01 \\
58841.3396868 & F350LP & 380 s & 16.54 $\pm$ 0.01 \\
58841.3459483 & F350LP & 380 s & 16.54 $\pm$ 0.01 \\
58841.3522099 & F350LP & 380 s & 16.53 $\pm$ 0.01 \\
58841.393402 & F350LP & 380 s & 16.52 $\pm$ 0.01 \\
58841.3996636 & F350LP & 380 s & 16.52 $\pm$ 0.01 \\
58841.4059252 & F350LP & 380 s & 16.52 $\pm$ 0.01 \\
58841.4121868 & F350LP & 380 s & 16.49 $\pm$ 0.01 \\
58841.4184483 & F350LP & 380 s & 16.51 $\pm$ 0.01 \\
58841.9895479 & F350LP & 380 s & 16.54 $\pm$ 0.01 \\
58841.9958095 & F350LP & 380 s & 16.54 $\pm$ 0.01 \\
58842.002071 & F350LP & 380 s & 16.5 $\pm$ 0.01 \\
58842.0083326 & F350LP & 380 s & 16.53 $\pm$ 0.01 \\
58842.0145942 & F350LP & 380 s & 16.54 $\pm$ 0.01 \\
58842.2544784 & F350LP & 380 s & 16.48 $\pm$ 0.01 \\
58842.26074 & F350LP & 380 s & 16.53 $\pm$ 0.01 \\
58842.2670016 & F350LP & 380 s & 16.53 $\pm$ 0.01 \\
58842.2732632 & F350LP & 380 s & 16.5 $\pm$ 0.01 \\
58842.2795247 & F350LP & 380 s & 16.53 $\pm$ 0.01 \\
58842.3207284 & F350LP & 380 s & 16.53 $\pm$ 0.01 \\
58842.32699 & F350LP & 380 s & 16.52 $\pm$ 0.01 \\
58842.3332516 & F350LP & 380 s & 16.52 $\pm$ 0.01 \\
58842.3395132 & F350LP & 380 s & 16.52 $\pm$ 0.01 \\
58842.3457747 & F350LP & 380 s & 16.53 $\pm$ 0.01 \\
58842.8515618 & F350LP & 380 s & 16.55 $\pm$ 0.01 \\
58842.8578233 & F350LP & 380 s & 16.55 $\pm$ 0.01 \\
58842.8640849 & F350LP & 380 s & 16.55 $\pm$ 0.01 \\
58842.8703465 & F350LP & 380 s & 16.52 $\pm$ 0.01 \\
58842.8766081 & F350LP & 380 s & 16.54 $\pm$ 0.01\\
\hline
\caption{Columns: (1) observation date correct for light travel time; (2) \textit{HST}/WFC3 Filter; (3) Exposure time (4) H magnitude with 1 $\sigma$ uncertainties}
\end{longtable}

\end{document}